\documentclass[11pt,twoside]{article}
\usepackage{FUSE2004}
\usepackage{natbib}
\usepackage{epsf}
\usepackage{psfig}
\usepackage{lscape}
\begin{document}
\title{Semi-empiric Radiative Transfer Modeling of FUSE Stellar Spectra}
\author{A. Lobel$^{1}$, E. H. Avrett$^{1}$, and J. P. Aufdenberg$^{2}$}
\affil{$^{1}$Harvard-Smithsonian Center for Astrophysics, 60 Garden Street,
  Cambridge 02138 MA, USA}
\affil{$^{2}$National Optical Astronomy Observatory, P.O. Box 26732, Tucson, 85726 AZ, USA}

\begin{abstract}
We present an overview of radiative transfer modeling efforts
to interpret spectra of a variety of stellar objects observed with {\it FUSE}. 
Detailed radiative transfer modeling of high ion emission line profiles 
of C\,{\sc iii} and O\,{\sc vi} observed in the far-UV spectrum, provides a powerful 
means to probe the thermal and dynamic properties of high-temperature 
plasmas in the atmospheres of stars. We model asymmetric 
emission lines of C\,{\sc iii} $\lambda$977 (and 
Mg\,{\sc ii} $h$ \& $k$) observed in spectra of luminous cool stars such as $\alpha$ Aqr, 
to infer the wind- and microturbulence velocity structures of the upper 
chromosphere. Semi-empiric radiative transfer models that include
transition region temperature conditions, are further developed based 
on detailed fits to O\,{\sc vi} resonance emission lines in the supergiant 
$\alpha$ Aqr, the classical Cepheid variable $\beta$ Dor, and to 
self-absorbed O\,{\sc vi} emission lines in the cataclysmic variable SW UMa.

We observe that the C\,{\sc iii} resonance line profile of $\alpha$ Aqr assumes 
a remarkable asymmetric shape, reminiscent of P Cygni type profiles observed in
hot luminous supergiants. The model calculations indicate outflow velocities 
above $\sim$140 $\rm km\,s^{-1}$ at kinetic temperatures of 65 kK and
higher. Based on detailed model fits to the narrow red-shifted and 
self-absorbed O\,{\sc vi}
emission lines of SW UMa we compute that the gas- and electron density exceed the 
density conditions of the upper solar transition region by about three orders 
of magnitude. We propose that the large gas density of $\rho$$\simeq$1.4 $10^{-11}$ 
$\rm gr\,cm^{-3}$ favors a region of warm dense plasma of 100 kK $\leq$$T_{\rm
  gas}$$\leq$ 300 kK that collapses onto the white dwarf with a mass accretion 
rate of 1$-$2 $10^{15}$ $\rm g\,s^{-1}$ above or between the accretion disk.
We discuss how detailed semi-empiric fits to emission lines 
observed with the high spectral resolution of {\it FUSE} can provide 
reliable constraints on the mass loss or mass accretion rates 
in these objects. The complete text of this conference presentation 
is available at cfa-www.harvard.edu/$\sim$alobel/conference.html
\end{abstract}

\begin{figure}[ht]
\plottwo{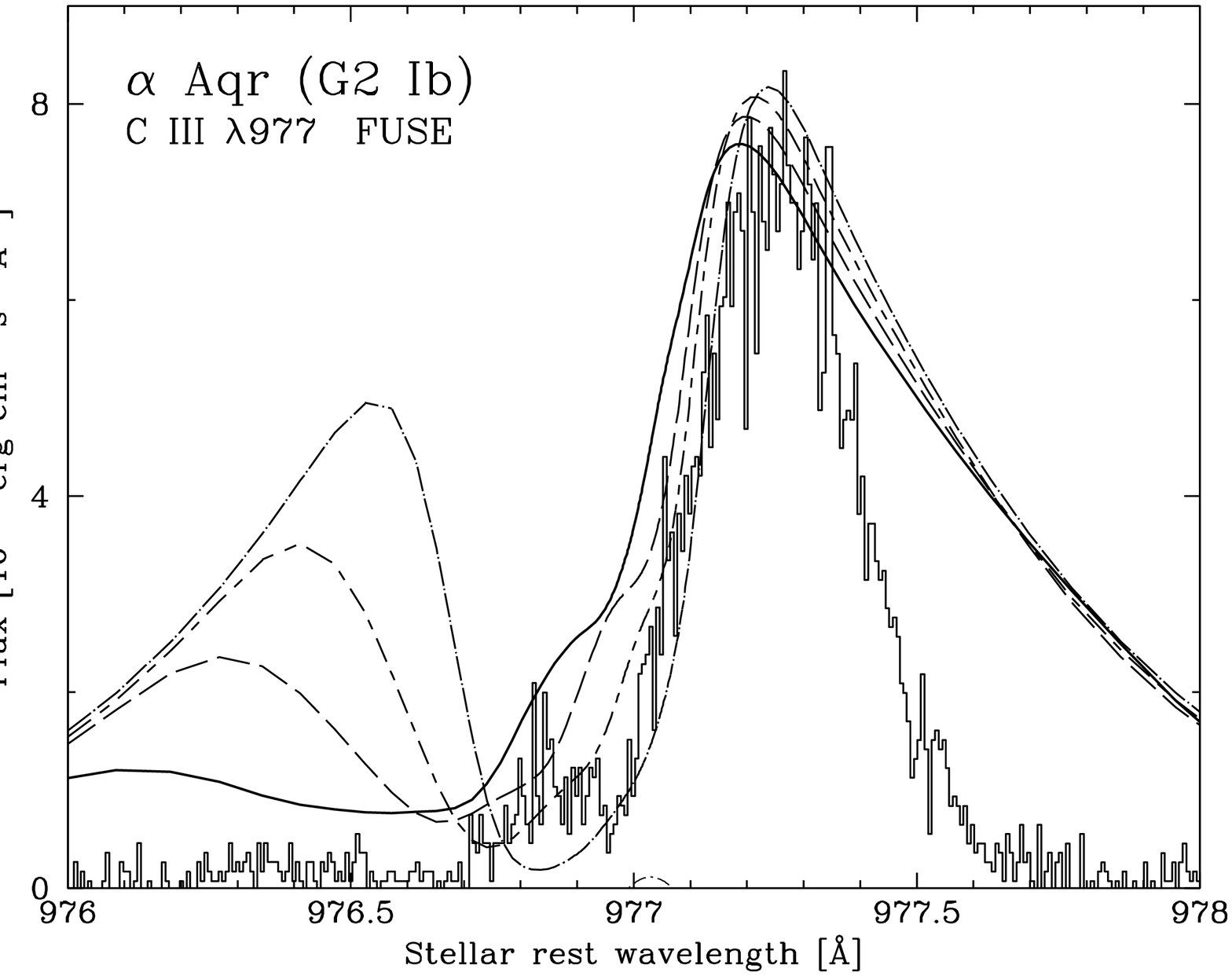}{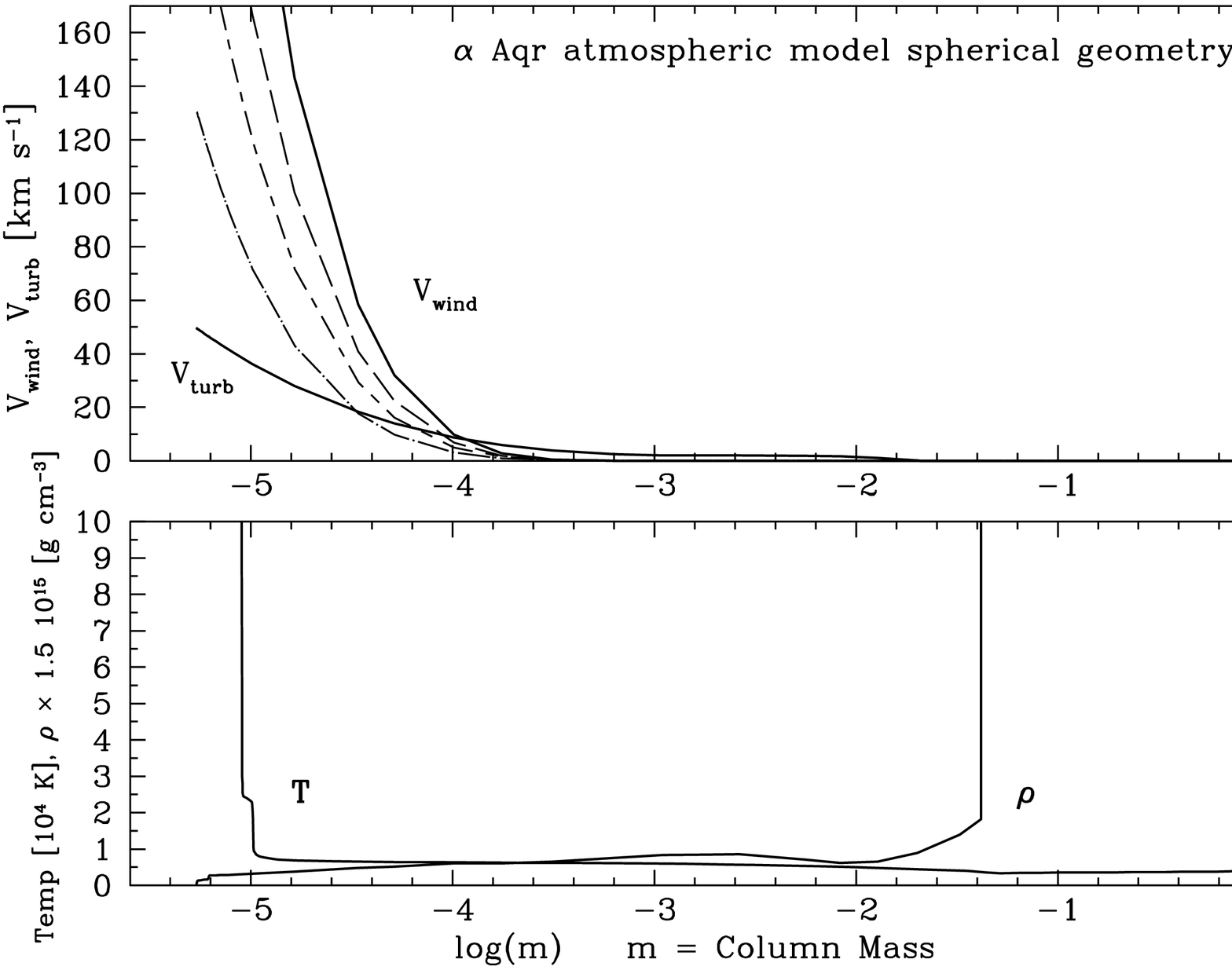}
\caption{Radiative transfer best fit ({\em solid line}) to C~{\sc iii}
  $\lambda$977 in $\alpha$ Aqr. }
\end{figure}

\section{{\it FUSE} Observations and Detailed Radiative Transfer Modeling}
We present a comparative study of transition region (TR) wind dynamics in 
$\alpha$ Aqr (G2 Ib), SW UMa (CV), $\beta$ Dor (F-G Ia-Iab) and the Sun,
to develop semi-empiric radiative transfer models of the thermal structure of 
the stellar chromosphere and TR, and to determine the velocity and electron density structures. 
{\it FUSE} spectra of the hybrid supergiant $\alpha$ Aqr have been
observed for Science Team Program P218 (Dupree et al. 2005).
The classical $\delta$-Cepheid variable 
$\beta$ Dor was observed in Aug. and Oct. 2003 for GI-D107 (PI A. Lobel),
while the dwarf nova SW UMa was observed in Nov. 2001 for
GI-B074. The left-hand panel of Fig. 1 
shows non-LTE radiative transfer fits to 
C~{\sc iii} $\lambda$977 through a semi-empiric model of the 
chromosphere and lower TR of $\alpha$ Aqr. The outward decrease of 
$\rho$ ({\em panels right}) produces a scattering core in the 
computed emission profile (Lobel \& Dupree 2000; 2001), which assumes 
an asymmetric shape due to opacity in a fast accelerating wind. 
Best multi-level atom model fits are obtained for 
$\rm V_{\rm wind}$$>$140~$\rm km\,s^{-1}$ in the lower TR 
($T$$\simeq$65 kK), which strongly scatters the blue emission line wing.
The model signals a supersonic optically thick warm wind in the 
outer atmosphere of this hybrid supergiant (Lobel \& Dupree 2002).

\begin{figure}[!b]
\plottwo{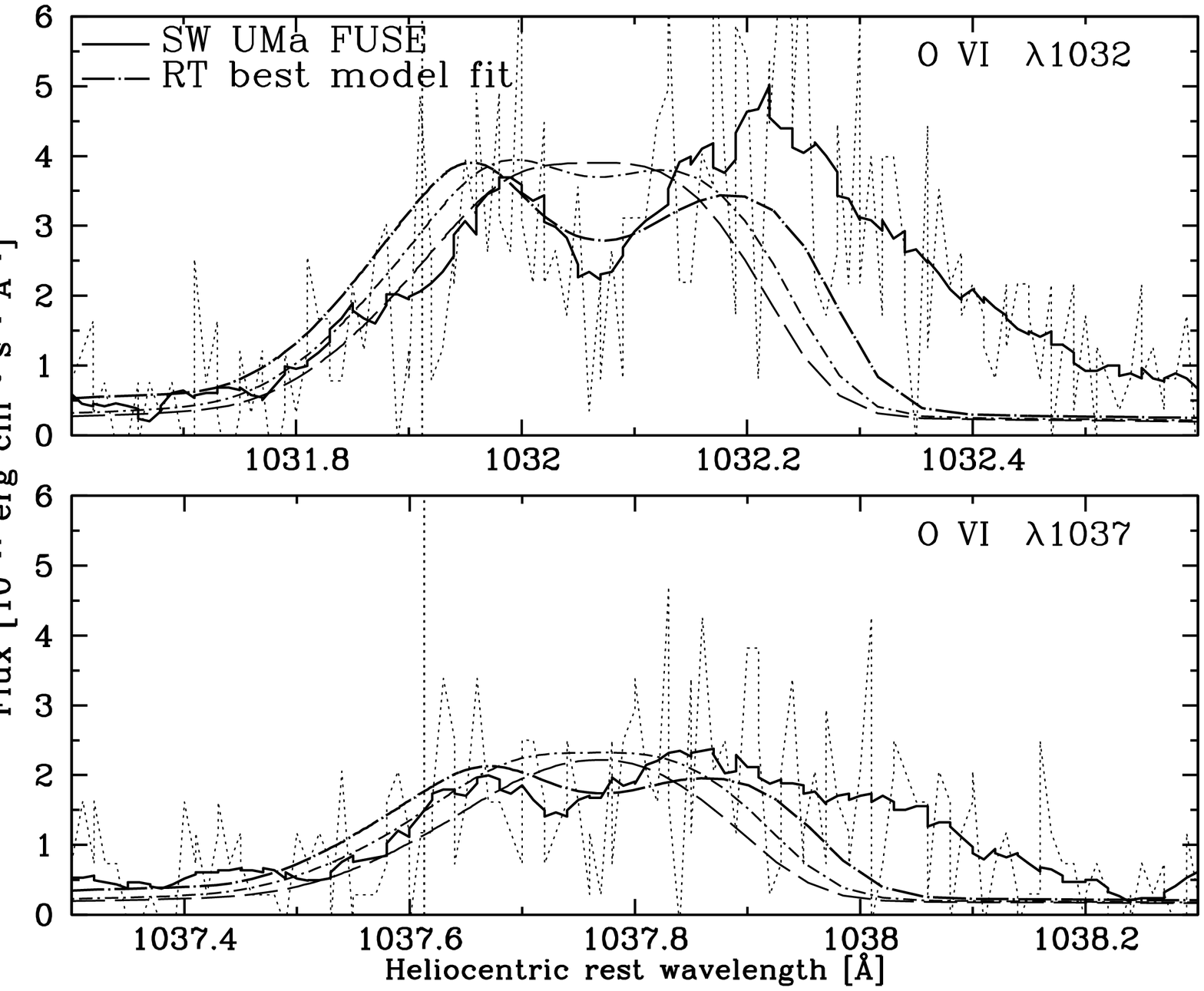}{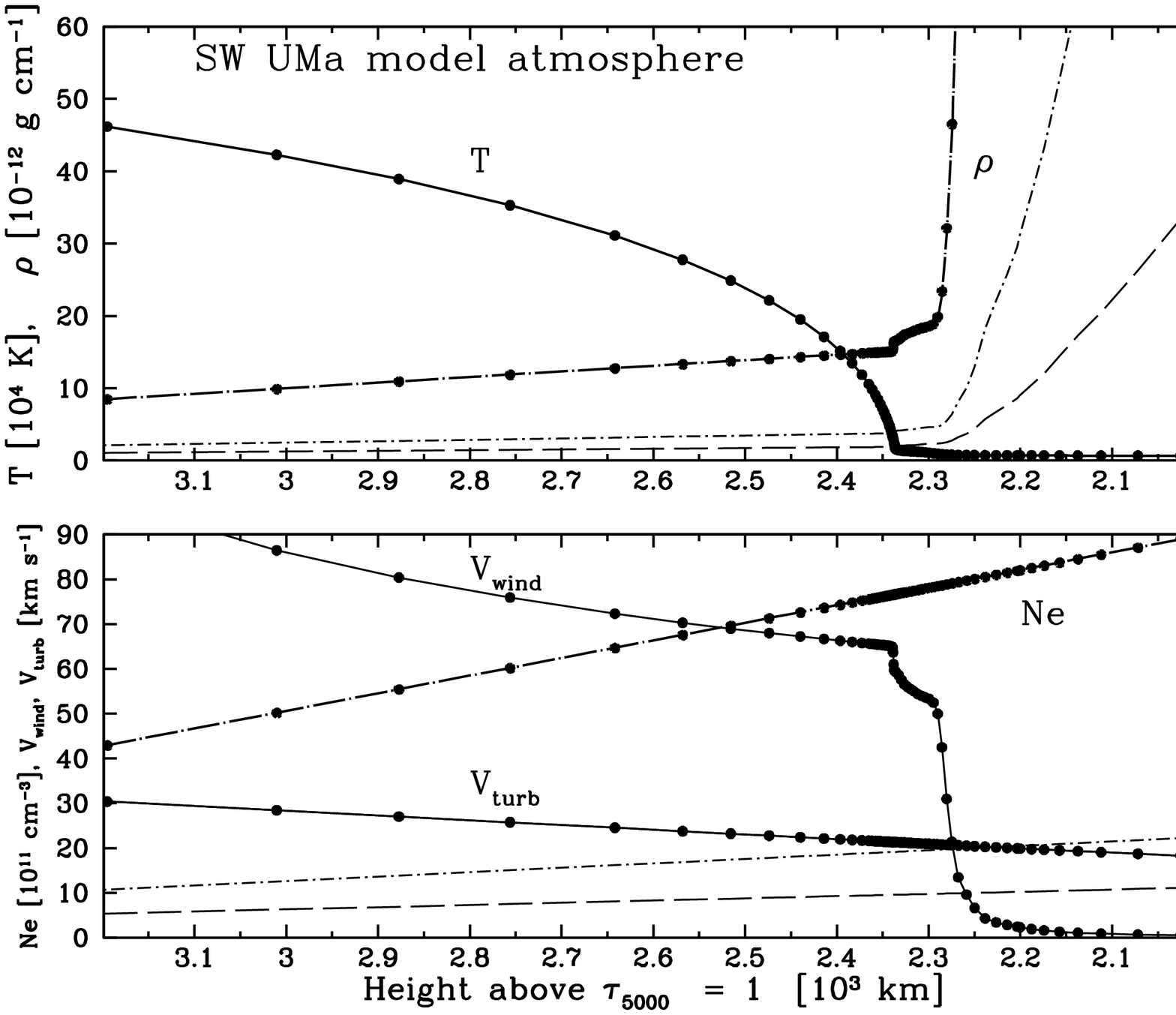}
\caption{Best fit ({\em long-dash dotted line}) to O~{\sc
    vi} $\lambda$1032 \& $\lambda$1037 in SW UMa.   }
\end{figure}

The left-hand panels of Fig. 2 show O~{\sc vi} lines in the 
cataclysmic variable SW UMa. The lines are remarkably far red-shifted 
with respect to heliocentric rest.
We suggest that the double-peaked narrow profiles result from 
self-absorbed emission line formation in an infalling region
near the white dwarf that is sufficiently optically thick at $T_{\rm gas}$
between 100 kK and 300 kK ({\em panels right}).
The depth of the O~{\sc vi} $\lambda$1032 central absorption line core 
can only be computed with electron densities $N_{e}$ at least 1100 
times larger compared to the solar TR values. 
The downflow velocity ({\em $V_{\rm wind}$}) in the model is computed from 
the conservation of total mass in spherical geometry with
$\dot{M}=4\,\pi\,\rho(r)\,V_{\rm wind}(r)\,(r+ R_{\rm WD})^{2}$, using a mass
accretion rate $\dot{M}$ of
1 $10^{15}$ $\rm g\,s^{-1}$, and $R_{\rm WD}$=0.01 $\rm R_{\odot}$. 
This $\dot{M}$-value provides the best fit to the wavelength position of 
the central self-absorption core where $V_{\rm wind}$=$+$70 to $+$65 $\rm km\,s^{-1}$
in the O~{\sc vi} lines formation region. 

\begin{figure}
\plottwo{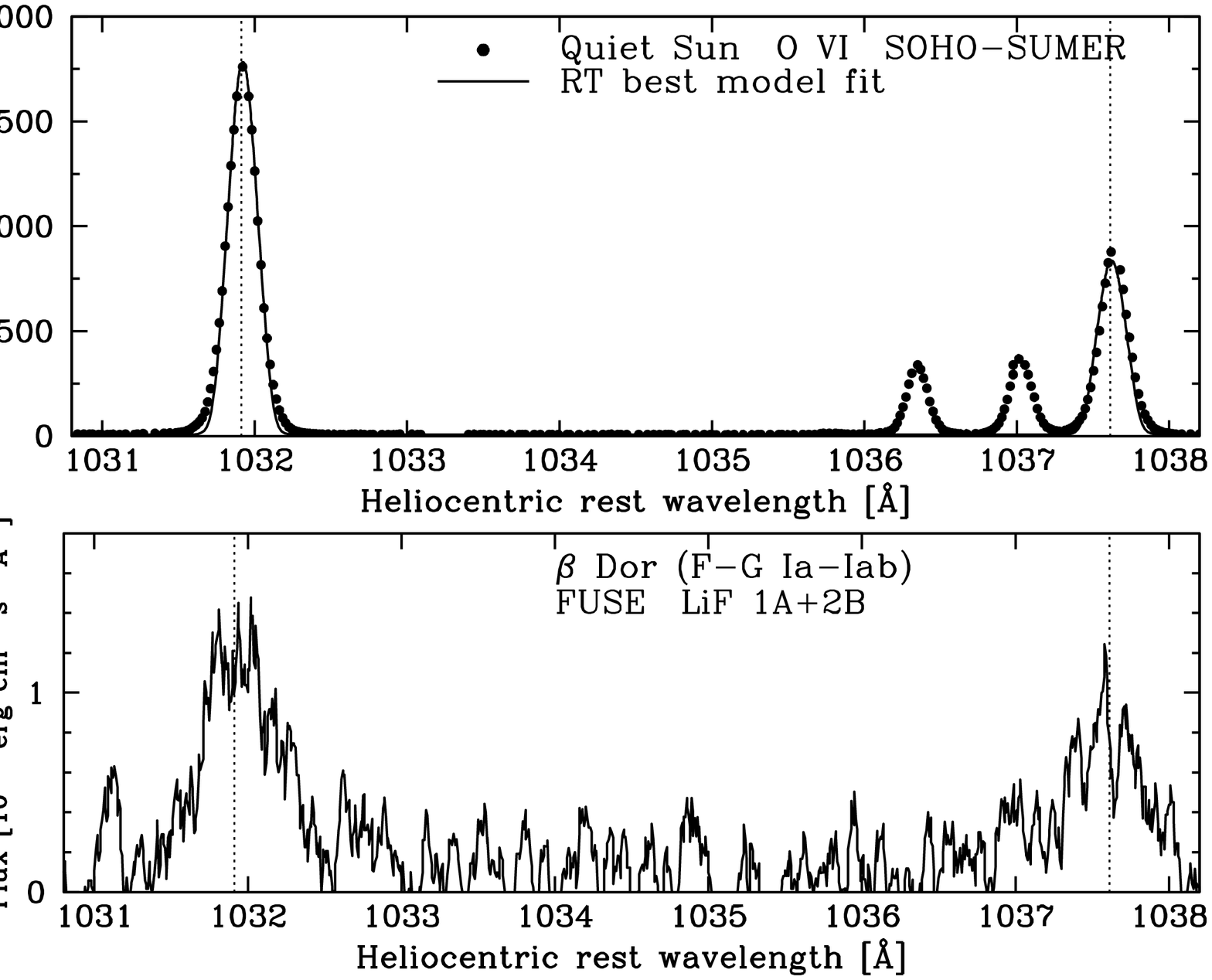}{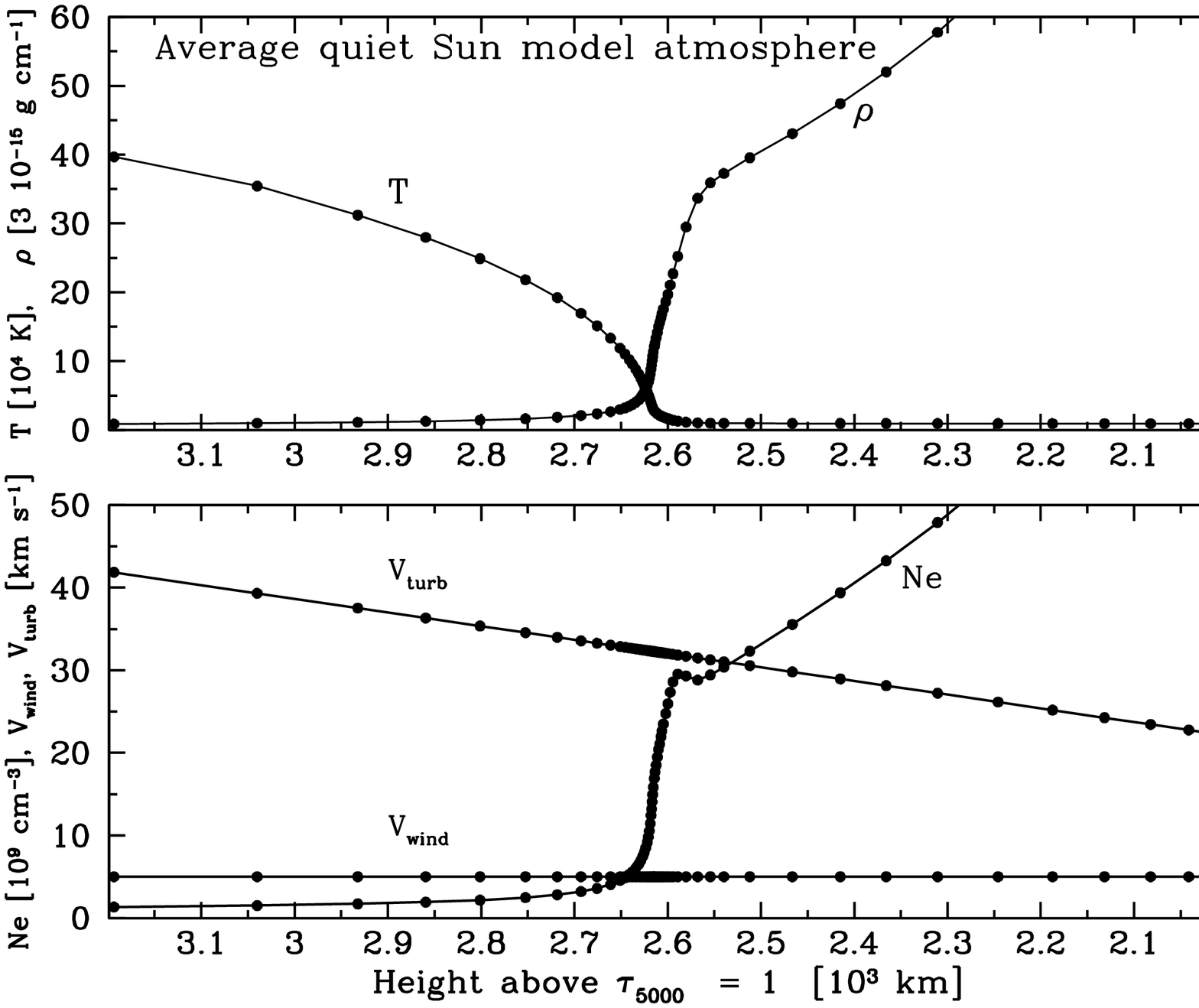}
\caption{Best fit to solar O~{\sc vi} ({\em upper left panel}) emission compared to
  $\beta$ Dor.}
\end{figure}

The upper left-hand panel of Fig. 3 compares the O~{\sc vi} lines we compute 
with an atmospheric model of the average quiet Sun 
with SOHO-SUMER observations ({\em solid dots}). 
The lines form over a rather small  
region of $\sim$10 model layers with 100 kK 
$\leq$ $T_{\rm gas}$ $\leq$ 300 kK, or $\sim$200 km into the solar TR
above the upper chromosphere ({\em panels right}).  A best fit to the FWHM and 
equivalent width of the lines is obtained with projected
microturbulence velocities ($ V_{\rm turb}$) increasing from 
at least $\sim$30 $\rm km\,s^{-1}$ to $\sim$39 $\rm km\,s^{-1}$ 
over this thermal range (Warren et al. 1997). 
A best fit to the core position of the O~{\sc vi} 
$\lambda$1032 line is computed with a mean downflow velocity of 
$+$5 $\rm km\,s^{-1}$. Preliminary atmosphere models
for $\beta$ Dor ({\em lower left panel}) require $ V_{\rm turb}$ 
exceeding 40 $\rm km\,s^{-1}$ in the O~{\sc vi} lines formation 
region to match the large line widths we also observe in $\alpha$ Aqr.  

\acknowledgments
Financial support for this research has been provided by 
NASA {\it FUSE} grants GI-D107 and GI-E068.

\end{document}